# Interfacial performance evolution of ceramics-in-polymer composite electrolyte in solid-state lithium metal batteries

Ao Cheng [1, †], Linlin Sun [2, †], Nicola Menga [3, †], Wanyou Yang [4], Xin Zhang [1, *]

1. School of Mechanical and Electrical Engineering, University of Electronic Science and Technology of China, Chengdu 611731, China

2. School of Mechanical Engineering, Chang'an University, Xi'an 710018, China

3. Department of Mechanics, Mathematics and Management, Polytechnic University of Bari, Via E. Orabona 4, 70125 Bari, Italy

4. School of Aeronautics and Astronautics, University of Electronic Science and Technology of China, Chengdu 611731, China

* Corresponding authors: Xin Zhang, zhangxin@uestc.edu.cn

**Abstract**：The incorporation of ceramics into polymers, forming solid composite electrolytes (SCEs) leads to enhanced electrical performance of all-solid-state lithium metal batteries. This is because the dispersed ceramics particles increase the ionic conductivity, while the polymer matrix leads to better contact performance between the electrolyte and the electrode. In this study, we present a model, based on Hybrid Elements Methods, for the time-dipendent Li metal and SCE rough interface mechanics, taking into account for the oxide (ceramics) inclusions (using the Equivalent Inclusion method), and the viscoelasticity of the matrix. We study the effect of LLTO particle size, weight concentration, and spatial distribution on the interface mechanical and electrical response. Moreover, considering the viscoelastic spectrum of a real PEO matrix, under a given stack pressure, we investigate the evolution over time of the mechanical and electrical performance of the interface. The presented theoretical/numerical model might be pivotal in tailoring the development of advanced Solid State Batteries with superior performance; indeed, we found that conditions in the SCE mixture which optimize both the contact resistivity and the interface stability in time.

# 1. INTRODUCTION

All-solid-state lithium metal batteries (ASSLMBs) are well-recognized for their high energy density, enhanced output voltage, extended cycle life, and minimal capacity degradation, making them promising candidates for electrochemical energy storage systems and electric vehicles, in contrast to the prevailing use of liquid-electrolyte lithium batteries [1-3]. One significant challenge the ASSLMBs faced with is the poor contact performance of the interface between the stiff solid-state electrolytes (SSEs, *e.g.* common oxides, sulfides) and the electrodes, which leads to high interfacial electric resistance, as well as components wear [4, 5]. The pioneering concept of a polymer/ceramic/polymer sandwich electrolyte (PCPSE) was first introduced by Goodenough and colleagues [6], marking significant improvements of the electrolytes-electrodes interface. While softer polymer coatings may enhance the interfacial contact fraction [7], they also lead to lower ionic conductivity than oxides, potentially undermining efforts to reduce overall electrical resistance.

A promising approach to address this trade-off is to incorporate oxide particles into polymers, forming solid composite electrolytes (SCEs) that combine high ionic conductivity with good contact performance with electrodes [8, 9]. In these SCEs, the migration of Li+ cations predominantly occurs through the motion of polymer chains, primarily within amorphous regions. Incorporating traditional additives such as plasticizers and nano-fillers into the polymer matrix can reduce crystallinity, thereby enhancing ionic conductivity by creating more amorphous regions that facilitate ion transport [10]. On the other hand, metal-organic frameworks (MOFs), when similarly incorporated into the polymer matrix, offer a different mechanism for enhancing the performance of SSEs. MOFs can provide additional ion-conducting pathways through their well-ordered porous structures, independent of the polymer matrix's crystallinity. Furthermore, MOFs can interact with the polymer chains and lithium ions, forming stable coordination sites that facilitate faster ion migration. The inclusion of MOFs not only improves ionic conductivity but also enhances the mechanical strength of the composite, making them a promising addition to solid composite electrolytes alongside traditional additives [11-17]. The fundamental mechanical challenge within oxide-polymer composite electrolytes is that the system is no longer homogeneous, as oxides represent elastic inclusions into the viscoelastic polymeric matrix [18-20]. This means that at the electrode-electrolyte interfaces in the ASSLMBs, a very complex contact behavior is expected due to the combined effects of viscoelasticity, inhomogeneity, and surface roughness. Those combined effects lead to significant complications to the theory and modeling development of solid-state batteries towards formulating interfacial contact improvement and electrolyte optimization strategies [21].

The determination of effective mechanical properties of inhomogeneous ceramics-in-polymer composite electrolyte systems is of paramount importance to the electrochemical-mechanical responses in battery performances [22-25], known as the inclusion problems, particularly considering the significant impact of the interaction of the reinforcing phases [26-28]. The Galerkin-vector method [29-31] and Green-function method [32, 33] have been recognized as efficient approaches for determining the closed-form solutions to inclusion problems with eigenstrains. Additionally, Eshelby's equivalent inclusion method (EIM) [34] provides an effective way to model the effect of generic inclusions, ellipsoidal [35, 36] or of arbitrary shape [37], as a perturbation to the elastic field in the homogeneous matrix. Furthermore, a modulus-perturbation approach has been recently employed to examine the indentation test of inhomogeneous materials [38, 39].

More importantly, most of existing studies on viscoelastic contact, as well as on the effect of inclusions, focus on single asperity contacts; on the contrary, electrolytes-electrodes contacts in ASSLMBs also present roughness, which behaves as a game changer. Indeed, the earlier studies on viscoelastic contact mechanics were focused on the indentation problems between smooth indenters and viscoelastic solids. Groundbreaking research by Lee and Radok [40] investigated the contact pressures evolution under monotonically increasing normal indentation with Hertzian indenters.

Following their paradigm, extended models were proposed by Hunter [41], with application to impact of spheres, and Graham [42], for a general non-monotonic continuously varying contact area. Fast-forward several decades to 2,010 when Greenwood [43] introduced a model to solve the contact problems of viscoelastic half-spaces indented by an axisymmetric indenter. Argatov [44] provided analytical solutions addressing the rebound characteristics in indentation issues of viscoelastic layers, while Menga *et al.* [45] investigated their frictional response in sliding conditions against wavy indenters. Chen *et al.* [46] introduced a robust semi-analytical approach for addressing indentation issues involving a rigid indenter and a viscoelastic half-space. This semi-analytical approach offers the capability to accommodate a broad range of relaxation times for linearly viscoelastic materials as well as arbitrary loading profiles. It can be extended to simulate contact between rough surfaces. A semi-analytical solution was also proposed by Menga *et al.* [47] for the contact between a wavy surface and a viscoelastic half-space, thus generalizing the well-known Westergaard solution to the viscoelastic case. Contact mechanics studies involving viscoelastic thin layers were also focused on rough interfaces [48, 49]. These firstly assumed frictionless motion between the contacting surfaces [50] and then, included the effect of Coulomb interfacial friction [51]. It was shown that, in the latter case, coupling occurs between in-plane and out-of-plane displacements and stresses, which might significantly alter the overall contact performance in terms of real contact area, surface stress concentration, and macroscopic resulting friction [52, 53]. Recently, the interplay between viscoelasticity and adhesion was also deeply investigated [54, 55], showing that cyclic (time-varying) deformations can increase the real contact area, as well as the peeling toughness of viscoelastic coating [56, 57].

On the other hand, semi-analytical approaches to the case of viscoelastic solids with inclusions usually refer only to smooth contacts (*i.e.*, spherical indenters), as in the case of Koumi *et al.* [58] dealing with a viscoelastic matrix with ellipsoidal inclusions under Hertzian indentation. Furthermore, Koumi *et al.* [59] also extended the proposed approach to solve the three-dimensional rolling/sliding contact problem with inhomogeneous viscoelastic bodies.

To formulate interfacial contact improvement and electrolyte optimization strategies of solid-state batteries, this paper develops a real viscoelastic rough contact model of inhomogeneous ceramics-in-polymer composite electrolyte systems, considering the combined effects of viscoelasticity, inhomogeneity, and interfacial roughness. This interfacial contact model introduces a novel approach based on the hybrid element method (HEM) for the mechanical response of inhomogeneous composite electrolytes. The HEM also encompasses a numerical equivalent inclusion method (EIM). By applying the fast Fourier transform (FFT) algorithms, the HEM efficiently resolves micromechanics issues in isotropic materials affected by inhomogeneities. The developed model provides solutions for interfacial resistance, contact stress, and subsurface stress under varying applied forces, loading time, and inhomogeneous conditions.

## 2. FORMULATION AND MODELING

In this paper, we focus on the interface contact between solid-state batteries and do not address how additives and MOFs influence lithium ion conduction within these batteries. **Figure 1** illustrates the time-dependent three-dimensional (3D) interfacial contact problem of the lithium metal anode pressed on a ceramics-in-polymer (*i.e.* ceramic particles mixed into the polymer matrix) solid composite electrolyte (SCE) under a normal load $P = p_0 A_0$, with $p_0$ and $A_0$ being the stack pressure and nominal contact area, respectively. The contacting surfaces always have a certain degree of roughness due to surface manufacturing and lithium deposition, so that the real contact area $A < A_0$.

The time-dependent gap $g(x, y, t)$ is the local separation between the metallic lithium anode and the SCE surfaces at time $t$ (notice, $z = 0$). For the problem at hand, $g(x, y, t)$ results from different contributions, as

$$g(x,y,t) = u_{z0}^S(x,y,t) + u_{z0}^L(x,y,t) + s_1(x,y,t) + s_2(x,y,t) - \delta_0(t), \qquad (1)$$

where $u_{z0}^S(x,y,t)$ and $u_{z0}^L(x,y,t)$ are the SCE and lithium anode surface displacement fields, respectively, $\delta_0(t)$ is the contact penetration, and $s_1(x,y,t)$ and $s_2(x,y,t)$ are the surface morphology (*i.e.*, roughness height) SCE and lithium metal, respectively. In this study, we consider the case of constant applied normal load $P$, which entails, in general, a time-varying penetration $\delta_0(t)$. Of course, the opposite scenario (constant $\delta_0$) can be achieved by opportunely controlling $P$ over time.

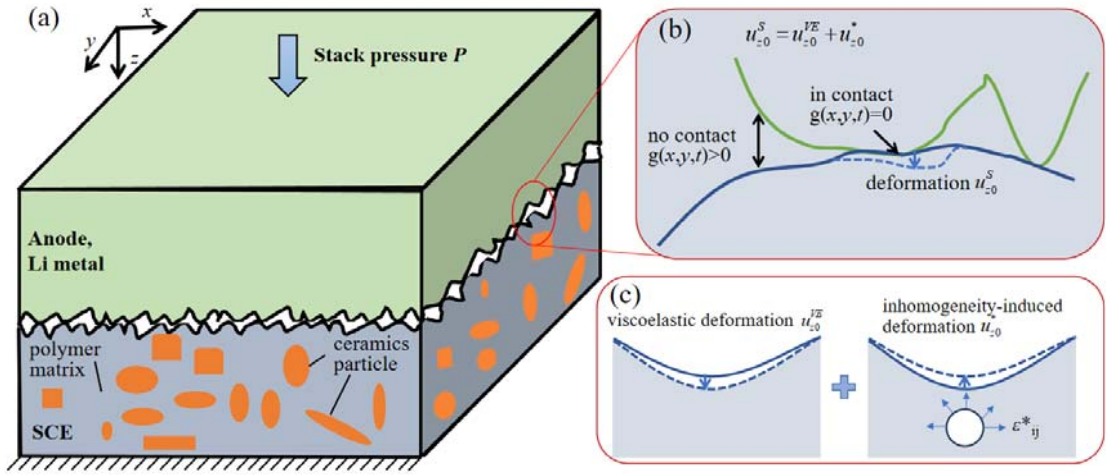

**Figure 1.** (a) Illustration of the time-dependent contact response of a lithium metal pressed on a ceramics-in-polymer solid composite electrolyte. (b) The gap and contact deformation in contact interface $u_{z0}^S(x,y,t)$. (c) The contact-induced viscoelastic deformation $u_{z0}^{VE}(x,y,t)$ and the inhomogeneity-induced deformation $u_{z0}^*(x,y,t)$.

**2.1 Displacement field of solid composite electrolytes**

Based on the theories of viscoelastic contact and the Eshelby's equivalent inclusion method (EIM) developed for materials embedded with inhomogeneities, the surface displacement $u_{z0}^S(x,y,t)$ of the solid composite electrolyte (SCE) is given by

$$u_{z0}^S(x,y,t) = u_{z0}^{VE}(x,y,t) + u_{z0}^*(x,y,t), \qquad (2)$$

where $u_{z0}^{VE}(x,y,t)$ is the surface viscoelastic displacement field associated to the polymeric matrix (assumed homogeneous), and $u_{z0}^*(x,y,t)$ is the displacement perturbation induced by the inhomogeneities.

In general, the linearly viscoelastic response is given by

$$\varepsilon(t) = \int_{-\infty}^{t} \Phi(t-\tau)\dot{\sigma}(\tau)\mathrm{d}\tau, \quad (3)$$

where $\varepsilon(t)$ and $\sigma(t)$ are the time-dependent strain and stress, respectively, and $\Phi(t)$ is the creep function (with $\Phi(t < 0) = 0$, for causality), which can be expressed as [60],

$$\begin{aligned}\Phi(t) &= \left[\frac{1}{E_0} - \int_0^\infty S(\tau)\exp(-t/\tau)\mathrm{d}\tau\right]H(t) \\ &= \left[\frac{1}{E_\infty} + \int_0^\infty S(\tau)\exp(-t/\tau)\mathrm{d}\tau\right]H(t),\end{aligned} \quad (4)$$

where $E_0$ and $E_\infty$ are the very low and very high frequency moduli, $S(\tau)$ is the continuously distributed creep spectrum at relaxation time $\tau$, and $H(t)$ is the Heaviside step function. Notably, in the elastic case $\Phi(t) = H(t)/E$, where $E$ is the Young's modulus.

Following the Green's function approach [59-62], the term $u_{z0}^{VE}(x,y,t)$ in Eq. (2) is given by

$$u_{z0}^{VE}(x,y,t) = \int_{-\infty}^{\infty}\int_{-\infty}^{\infty}\int_0^t G_{VE}(x-x', y-y', t-t')\frac{\partial p(x',y',t')}{\partial t'}\mathrm{d}x'\mathrm{d}y'\mathrm{d}t', \quad (5)$$

where $G_{VE}(x,y,t)$ is the viscoelastic Green's function for the (homogeneous) polymer matrix of the SCEs, and $p$ is the contact pressure. Moreover, invoking the elastic-viscoelastic correspondence principle [59-62] and using Eq. (4), we have $G_{VE}(x,y,t) = \frac{(1-v^2)\Phi(t)}{\pi}\frac{1}{\sqrt{x^2+y^2}}$.

Using fast Fourier transform (FFT) algorithm allows to achieve efficient numerical solution of Eq. (5) in the spatial frequency domain [63, 64]. In this study, we rely on the continuous convolution-fast Fourier transform (CC-FFT) algorithm, as better detailed in Refs. [65, 66], which is better suited for areal contact conditions. Using spatial CC-FFT and time discretization (j-index), Eq. (5) can be rewritten as,

$$u_{z0}^{VE}(x,y,N_t \Delta t) = \sum_{j=0}^{N_t} \left\{ \text{IFFT}[\hat{C}_{u_{z0}}^{VE}(m,n,(N_t - j)\Delta t) \cdot (\hat{p}_j - \hat{p}_{j-1})] \right\} \quad (6)$$

where the symbol $\hat{\cdot}$ indicates the CC-FFT operation (with $m,n$ being the spatial frequencies), $\hat{C}_{z0}^{VE}$ are the time-dependent Fourier transformed influence coefficients (ICs), and $N_t$ is the number of time steps considered (i.e., $t = \Delta t\, N_t$).

The perturbation displacement fields induced by elastic inclusions within the viscoelastic matrix in solid composite electrolyte, i.e. the term $u_{z0}^*(x,y,t)$ in Eq. (2), is calculated by relying on Eshelby's equivalent inclusion method (EIM) [34], opportunely modified to deal with semi-infinite solids [67, 68] based on Chiu's mirroring decomposition. Following Zhou et al. [69], within each elastic inclusion, the consistency condition of EIM on the strain field $\varepsilon$ is enforced as,

$$C_{ijkl}^*(\varepsilon_{kl}^0 + \varepsilon_{kl}) = C_{ijkl}(\varepsilon_{kl}^0 + \varepsilon_{kl} - \varepsilon_{kl}^*) \quad (7)$$

where $C_{ijkl}^*$ and $C_{ijkl}$ are the stiffness constants of the elastic inclusion and viscoelastic matrix, respectively, $\varepsilon_{kl}^0$ is the strain induced by a generic surface pressure distribution in a perfectly homogeneous solid (with stiffness tensor $C_{ijkl}$), $\varepsilon_{kl}$ is the perturbation strain due to the presence of the inclusion, and $\varepsilon_{kl}^*$ is the equivalent eigenstrain. Using the numerical procedure outlined in [29, 69-71], given the size and shape of the inclusions, the eigenstrain $\varepsilon_{kl}^*$ and the perturbation displacements field $u_{z0}^*$ can be calculated from $C_{ijkl}^*$ and $C_{ijkl}$ which depend, respectively, on the elastic modulus of the inclusion and on the equivalent time-dependent viscoelastic stiffness $1/\Phi(t)$.

### 2.3 Contact model numerical implementation

The analytical treatment of eigenstrain-induced elastic fields in viscoelastic substrates presents a marked increase in complexity relative to their elastic counterparts. Viscoelastic materials exhibit a distinct time-dependent strain-stress relationship, which becomes particularly intricate when inhomogeneities are introduced (e.g., material composition variations or structural imperfections). The resulting enhanced complexity necessitate an advanced numerical implementation for the contact model described so far.

Indeed, given the (time-varying) contact region $\Gamma_t$ at time $t$, the contact conditions for the instantaneous pressure $p(x, y, t)$ and gap $g(x, y, t)$ distributions can be written as

$$p(x,y,t) > 0 \text{ and } g(x,y,t) = 0 \text{ in } \Gamma_t,$$
$$p(x,y,t) = 0 \text{ and } g(x,y,t) > 0 \text{ out of } \Gamma_t,$$
$$\int_{\Gamma_t} p(x,y,t)\, dxdy = P, \tag{8}$$

and are enforced, together with the Eq. (1), by means of conjugate gradient method minimization of pressure residuals [66].

The overall solution algorithm for the contact is given by the following steps (see also the flowchart in **Figure 2**):

1) The surface roughness of SCEs and Li metal, the stack pressure, the viscoelastic properties of polymer matrix, the distribution, shape, and elastic properties of ceramics particles, are given as input parameters.

2) The viscoelastic contact displacements, deformations, and stresses of the (homogeneous) polymer matrix are calculated by means of conjugate gradient method (CGM) and the CC-FFT.

3) The inhomogeneity-induced perturbation displacements, deformations, and stresses are calculated by the equivalent inclusion method (EIM) and the CC-FFT.

4) The surface contact geometry is modified by incorporating the effect of SCEs inhomogeneity, and the contact pressure distribution is updated.

5) The Li elastic displacements are then computed based on the updated contact

pressure.

6) This process is repeated until the difference between the eigen-displacements of two consecutive iterations falls below the prescribed error.

7) The stress, contact fraction, and interface resistance are calculated.

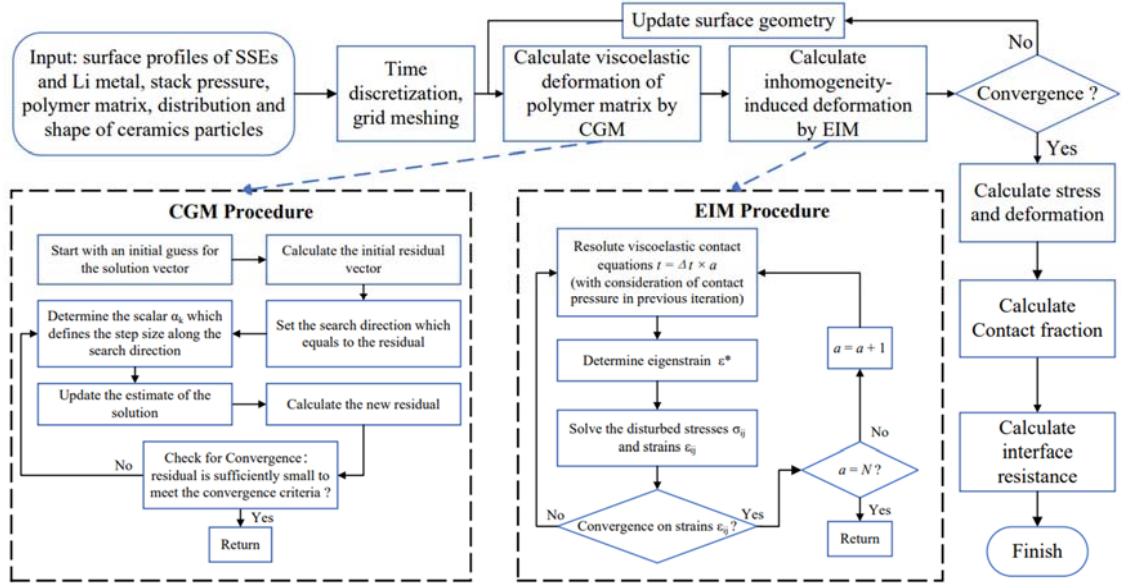

**Figure 2.** Flowchart of the algorithm for the coupling between the viscoelastic contact problem and the inhomogeneity problem of the interface in solid-state batteries

## 3. RESULTS AND DISCUSSION

### 3.1 Model validation

We firstly present the results for a single inclusion into an elastic matrix, and we validate our prediction by comparison with those by Zhou *et al.* [71]. The computational domain is discretized into a uniform grid system consisting of 128 × 128 × 128 grids.

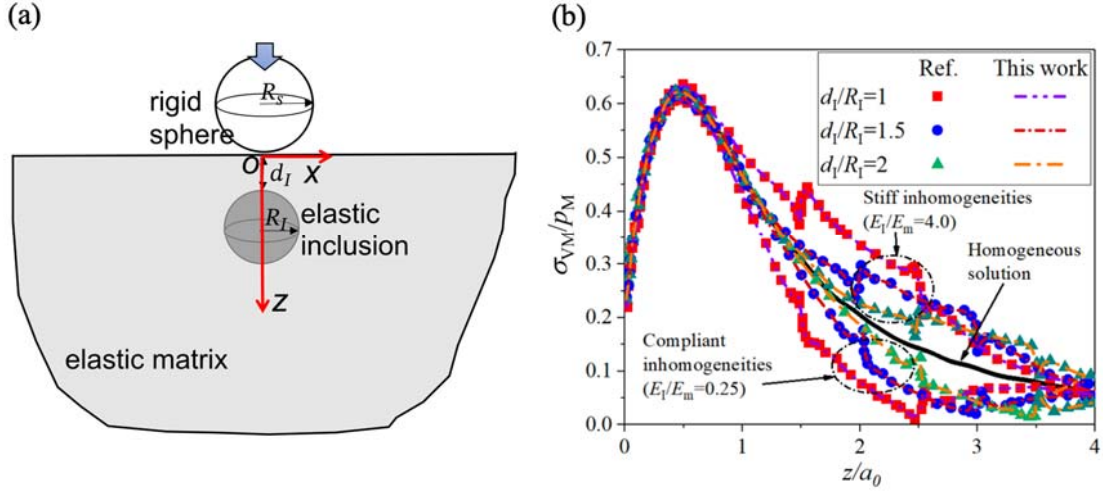

**Figure 3.** (a) Schematic of the contact between a rigid spherical indenter of radius $R_S$ and an elastic matrix with a spherical elastic inclusion of radius $R_S$ localized at a depth $d_I$. (b) Comparison of the nomalized von Mises stress distributions $\sigma_{VM}/p_M$ along the z-direction obtained from the presented model and those by Zhou *et al.* [71], where $R_I = R_S$, $d_I/R_I = 1, 1.5, 2$. Notice, $p_M$ is the peak Hertzian pressure.

The model, illustrated in **Figure 3**(a), involves a rigid sphere with a radius of $R_S$ in contact with an elastic matrix containing an elastic and spherical inhomogeneity. The radius of the spherical inhomogeneity is set to $R_I = R_S$. Its depth, $d_I$, is defined as the distance from the top surface of the spherical inhomogeneity to the matrix surface, with $p_M$ and $a_0$ being the peak pressure and contact radius of the equivalent Hertzian solution. The Young's moduli of the inhomogeneity and the matrix are denoted as $E_I$ and $E_m$, respectively. We consider both the cases of softer ($E_I/E_m = 0.25$) and stiffer ($E_I/E_m = 4.0$) inhomogeneities, compared to the matrix. In **Figure 3**(b), we show the von Mises stress distribution along the *z*-direction, corresponding to different inhomogeneity depth, which are in strong agreement with the predictions by Zhou *et al.* [71].

The second validation we present refers to the case of steady rolling (or frictionless sliding) of a rigid sphere against a viscoelastic half-space containing a spherical elastic inclusion, as shown in **Figure 4**(a), under given contact penetration $\delta$ instantaneously applied at time $t = 0$. The same problem was addressed by Koumi *et al.* [59] employing Eshelby's formalism iteratively throughout the temporal discretization process, thus investigating the inhomogeneity's impact on various factors, including contact pressure distribution, sub-surface stress, rolling friction, and resultant torque necessary to sustain the motion. For an accurate comparison, we set the same parameters as in Koumi *et al.* [59]; therefore, in what follows, the viscoelastic creep function for the matrix is defined as

$$J(t) = \left[\frac{1}{E_0} - \frac{1}{E_1}exp\left(-\frac{t}{\tau}\right)\right] = \left[\frac{1}{E_\infty} + \frac{1}{E_1}(1 - exp\left(-\frac{t}{\tau}\right))\right] \quad (9)$$

with $E_\infty = 10.036$ MPa, $E_\infty / E_0 = 10$, and $\tau = 0.01s$. The simulation time ranges in $(0, 2\tau)$, with 60 time steps.

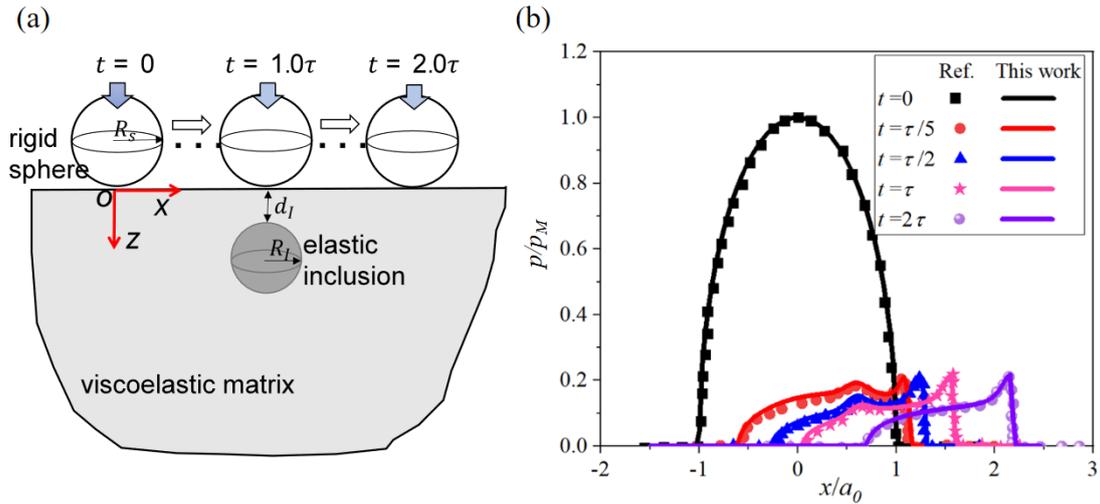

**Figure 4.** (a) Illustration of a spherical indenter sliding on a viscoelastic half-space containing a spherical elastic inclusion. (b) Comparison for distribution of normalized contact pressure $p/p_M$ of the viscoelastic half-space containing a spherical inhomogeneity with $\gamma = E^I/E_\infty = 1$ and $R_I = 0.25a_0$. The prescribed normal penetration is $\delta = 0.1$ mm and the dimensionless sliding speed is $v\tau/a_0 = 0.6$, with $a_0$ being the instantaneous contact area radius at $t = 0$. The radius of the spherical indenter is $R_S = 10$ mm. The Poisson's ratio of the inhomogeneity is identical to that of the viscoelastic half-space, $v^I = v^M = 0.3$. Notice, $p_M$ is the peak Hertzian pressure with elastic modulus $E_\infty$.

**Figure 4**(b) shows the comparison between the normalized pressure distributions $p/p_M$ at different time predicted with the present model and by Koumi *et al.* [59]. Specifically, calculations assume rigid sphere motion at constant velocity $v = 0.6\ a_0/\tau$ under a fixed normal displacement $\delta = 0.1$ mm are calculated by a degenerated form of the presented model. Again, very good agreement is found even in the case of single asperity viscoelastic contacts with elastic inhomogeneity, with the peak pressure localized at the contact leading edge as prescribed for viscoelastic rolling (or frictionless sliding) contacts [39,47].

**3.2 Viscoelastic contact with rough surfaces**

3.2.1 Contacts of viscoelastic homogeneous matrix

In this section, we investigate the contact between a rigid wavy indenter and a viscoelastic homogeneous matrix (i.e., with no inclusions), as shown in **Figure 5**(a). We focus on the exemplar case of a Li metal and the solid-state polymer electrolyte interface [72], with nominal contact area $A_0 = L_c \times L_c$, where $L_c = 400$ μm, and the mesh points are set to $110 \times 60 \times 60$ along the *x*, *y* and *z* directions. indicates the fundamental wavelength of the interface, as descried in. The. To simplify the calculation, as shown in **Figure 5**(b), we assume a square roughness on the Li metal surface with fundamental wavelength $\lambda_0 = 100$ μm; therefore, in Eq. (1) we have $s_2 = 6.05 cos(2\pi x/\lambda_0)cos(2\pi y/\lambda_0)$.

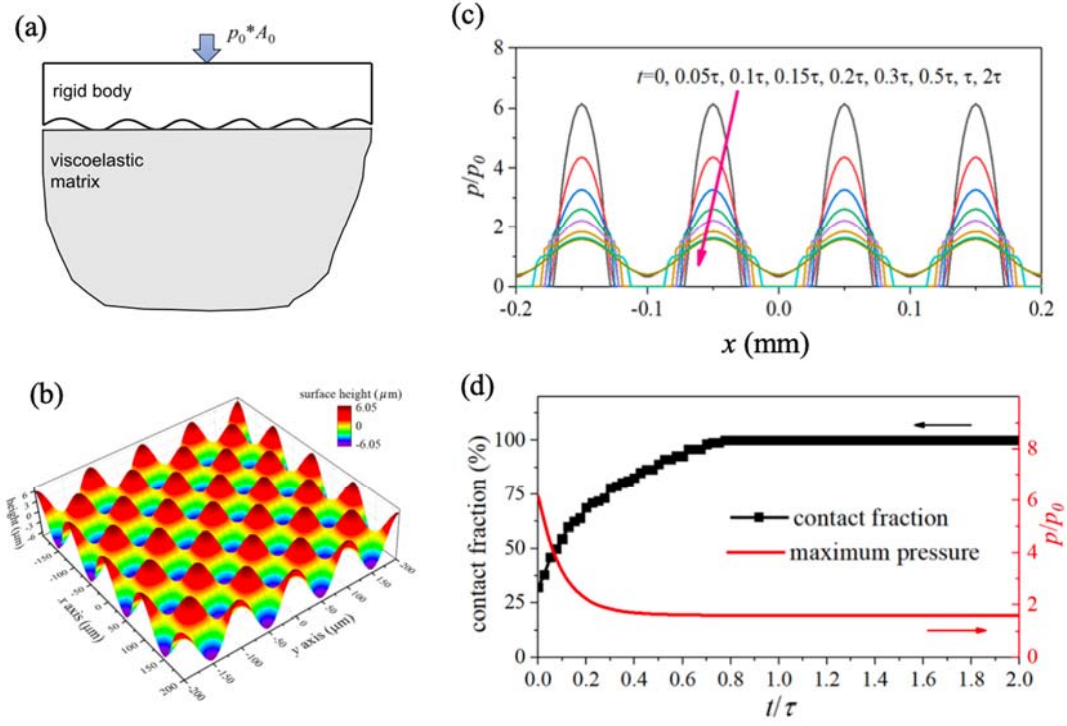

**Figure 5.** (a) The 3D rough contact model with a viscoelastic matrix. (b) The topography map of the Li metal surface with a typical cosine function. (c) The distribution of normalized surface pressure along the x-axis (*i.e.*, for $y = 0$) at different times. (d) The evolutions of the surface contact fraction, and the maximum surface pressure with time.

Moreover, the stack pressure $p_0 = 0.5$ MPa is instantaneously applied at time $t = 0$, and the focus on the time evolution of the normalized pressure distributions $p/p_0$ (**Figure 5**I) and contact fraction (**Figure 5**(d)). Initially, at $t = 0$, when the load is applied, viscoelastic polymer behaves almost purely elastically, with elastic modulus equal to the very-high frequency viscoelastic one $E_\infty$. Therefore, the contact spots are small and localized on the roughness summits, with high pressure peaks. As the contacting time increases, viscoelastic creep occurs, leading to contact softening; indeed, **Figure 5**(d) shows a decrease in contact peak pressure and an increase in the contact area. For $t \gg \tau$, the elastic behavior with elastic modulus equal to the very-low frequency viscoelastic one $E_0$ is asymptotically approached. Notably, at $t = 0.775\,\tau$, full contact conditions occur, with $A_c = A_0$, thus clearly indicating that the effective contact area fraction can significantly benefit from using soft polymer electrolytes, with improved (tailored) interface performance.

3.2.2 Contacts of viscoelastic matrix with spherical elastic inclusions

The primary goal of applying coatings on electrical interfaces is to enhance the real contact area, thereby reducing resistance. However, applying polymeric coatings also presents the drawback of lower ionic conductivity compared to the coated substrate (*i.e.*, the oxide solid-state electrolyte), which might even lead to worse performance. To overcome this issue, oxide particulates are dispersed within the polymer matrix, yielding solid composite electrolyte coatings. This necessitates the development of a real rough contact model that accounts for the combined effects of viscoelasticity and inhomogeneity.

In this section, we firstly consider a viscoelastic polymeric matrix with an elastic spherical inclusion, with $E^I = 5E_\infty$ and $\nu^I = \nu^M = 0.3$. As shown in **Figure 6**(a), the depth $d_I$ and radius $R_I$ of the inclusion are set at 30 μm. Our simulations span a time domain of $(0, \tau)$, divided into 50 time steps, with $p_0$ and $L_c$ being the same as in the previous section. Again, the stack pressure $p_0$ is instantaneously applied at time $t = 0$. **Figure 6**(b) displays the time-dependent evolution of the contact fraction and peak surface pressure, which reach steady values for $t \gtrsim 0.78\tau$. This clearly indicates that the effect of elastic inclusion on the viscoelastic creep of the interface is negligible. **Figures 6**(c) and (d) show the surface pressure along the *x*-axis and the stress distribution with the material along the *z*-axis (in the *x*O*z* section) at different time instants. The stress profiles reveal dual peak stress points in the z-direction, underscoring the influence of particle distribution on stress dynamics. Comparing **Figure 5**(c) to **Figure 6**(c), clearly highlights that the pressure distribution is no longer periodic over $\lambda_0$, as the presence of inclusions leads to a higher peak pressure on the asperity right above the inclusion.

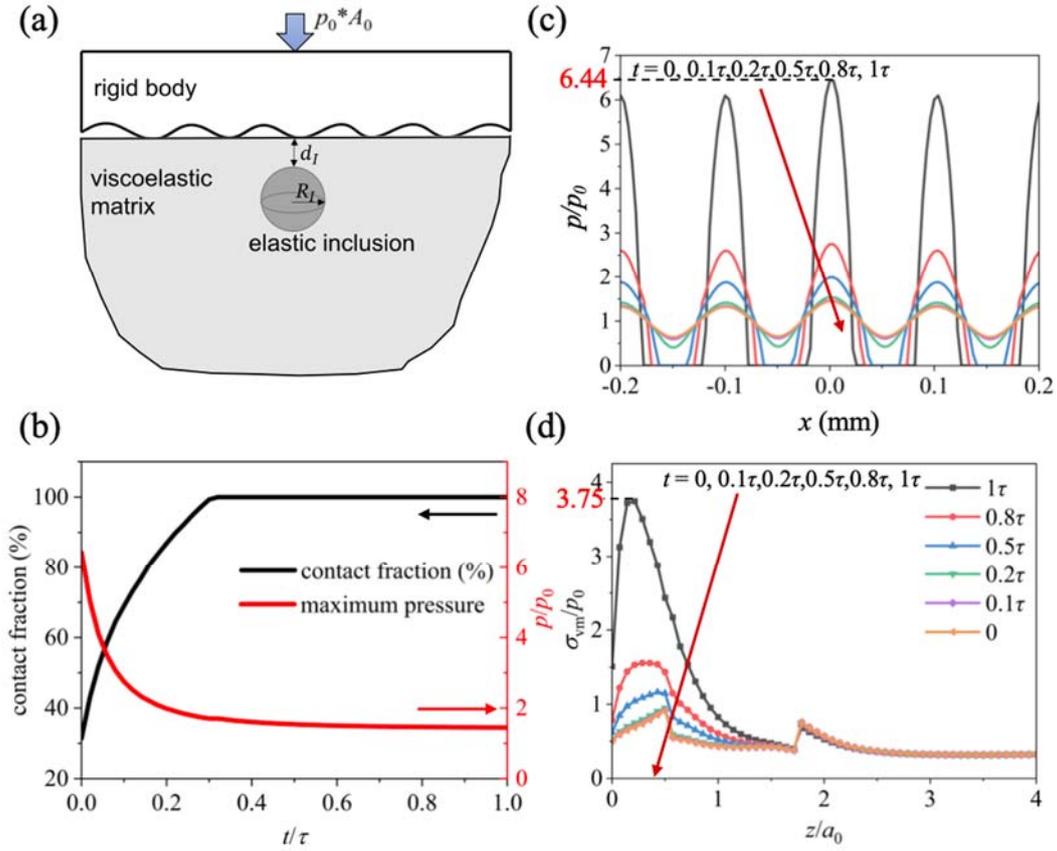

**Figure 6.** (a) A 3D rough surface contact model with viscoelastic matrix containing one elastic inclusion localized at $(x, y, z) = (0, 0, d_I)$, with elastic modulus $E_I = 5E_\infty$ and $d_I = R_I = 30$ μm. (b) The contact fraction and peak surface pressure as functions of normalized time. (c) The surface pressure distribution along the $x$-axis (*i.e.*, for $y = 0$) at different time instants. (d) The von Mises stress distributions along the $z$-axis (*i.e.*, in correspondence of the inclusion, where $(x, y) = (0,0)$) at different time instants ($a_0 = 50$μm is the instantaneous contact radius at time $t = 0$ for the homogeneous viscoelastic case).

**Figures 7** and **8** present contour maps of internal von Mises stress and surface pressure, respectively, which facilitate a deeper understanding of system evolution. Initially, both stress and surface pressure reach their highest values, while they reduce as $t$ increases and viscoelastic creep occurs. After $t = 0.8\tau$, in the contour maps of von Mises stress (see **Figure 7**), the effect of the inclusion is negligible.

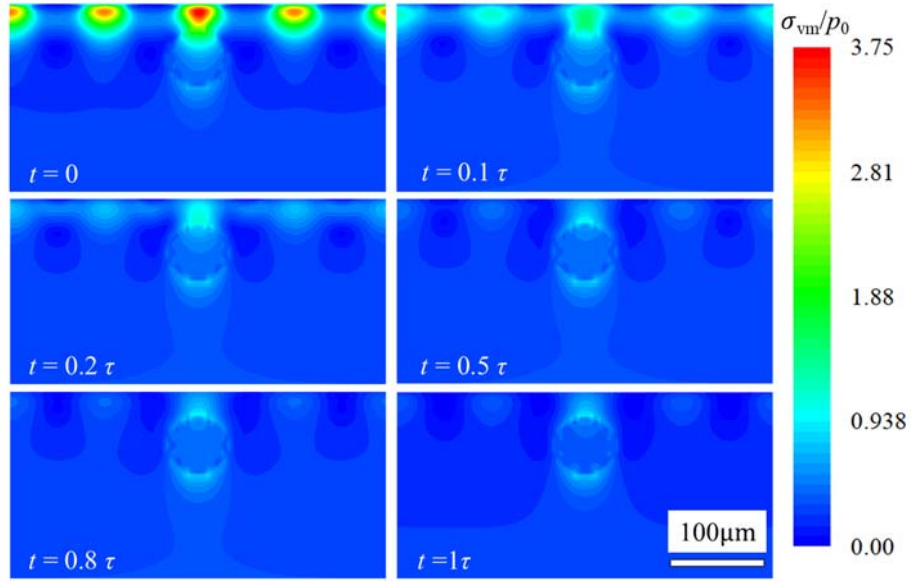

**Figure 7.** Contour maps of normalized von Mises stress at different time instants for a system with an inclusion under a stack pressure of $p_0$.

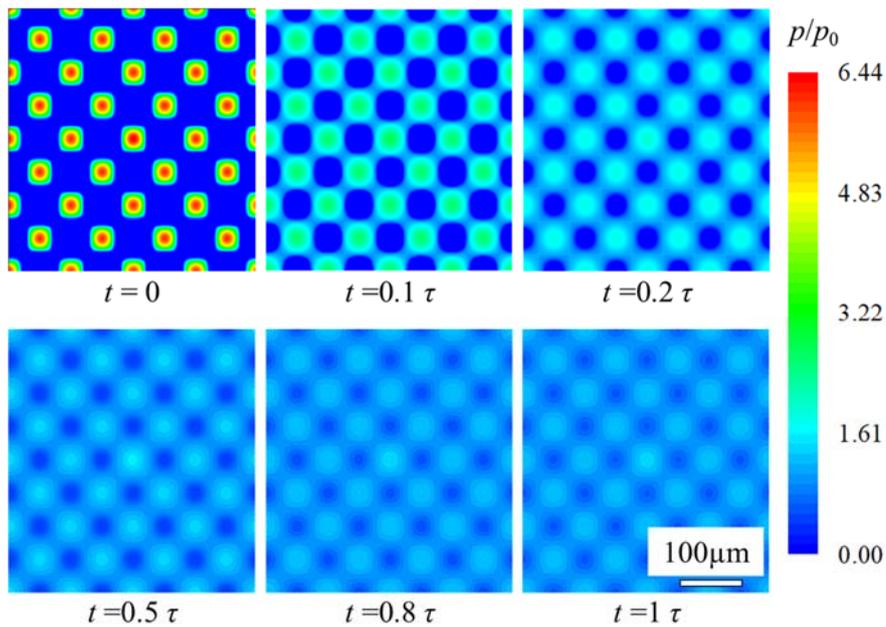

**Figure 8.** Contour maps of normalized surface pressure at different time instants for a system with an inclusion under a stack pressure of $p_0$.

As a second exemplar investigation, as shown in **Figure 9**(a), this time we consider a viscoelastic polymeric matrix with more than one elastic inclusion, namely in the number of three, equispaced by 65 μm. All the other parameters retain the same values as in the previous case.

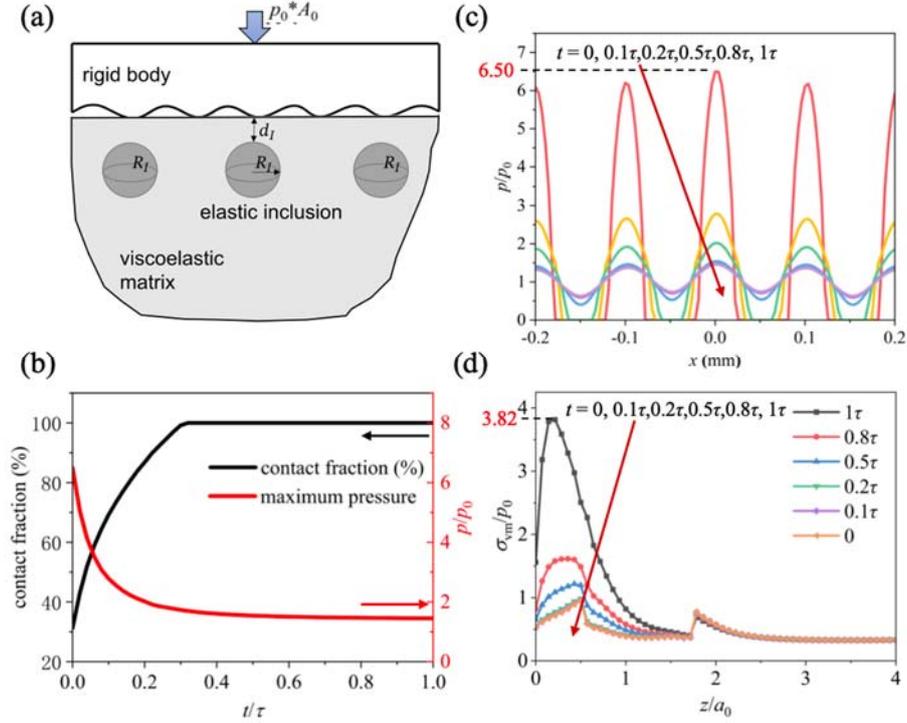

**Figure 9.** (a) A 3D rough surface contact model with a viscoelastic matrix containing three spherical elastic inclusions along the *x*-axis (*i.e.*, for $y = 0$) with elastic modulus $E_I = 5E_\infty$ and $d_I = R_I = 30$ μm. The distance between the inclusions' centers is 65 μm. (b) The contact fraction and peak surface pressure as functions of normalized time. (c) The surface pressure distribution along the *x*-axis (*i.e.*, for $y = 0$) at different time instants. (d) The von Mises stress distributions along the *z*-axis (*i.e.*, in correspondence of the inclusion, where $(x, y) = (0,0)$) at different time instants ($a_0 = 50$μm is the instantaneous contact radius at time $t = 0$ for the homogeneous viscoelastic case).

Comparing the two scenarios (*i.e.*, one to three inclusions cases), as depicted in **Figure 6**(b), the contact response shows a similar trend, with steady macroscopic behavior for $t \gtrsim 0.3\tau$ in both cases (see **Figures 6**(b) and **9**(b)). However, a closer look reveals that at $t = 0.28\tau$, the contact ratio with one inclusion is 99.32%, whereas it increases to 99.17% with three inclusions. This indicates that an increase in the number of inclusions slightly reduces the overall interface stiffness. Also, the other quantities in **Figures 9**(c) and (d) reveal slight variations from one to three inclusions cases (*i.e.*, the maximum subsurface von Mises stress and the peak surface pressure). Such a tiny quantitative dependence of the results on the specific number of inclusions is ascribable to their very local effect on the elastic fields in the SCE, so that they almost do not interact with each other.

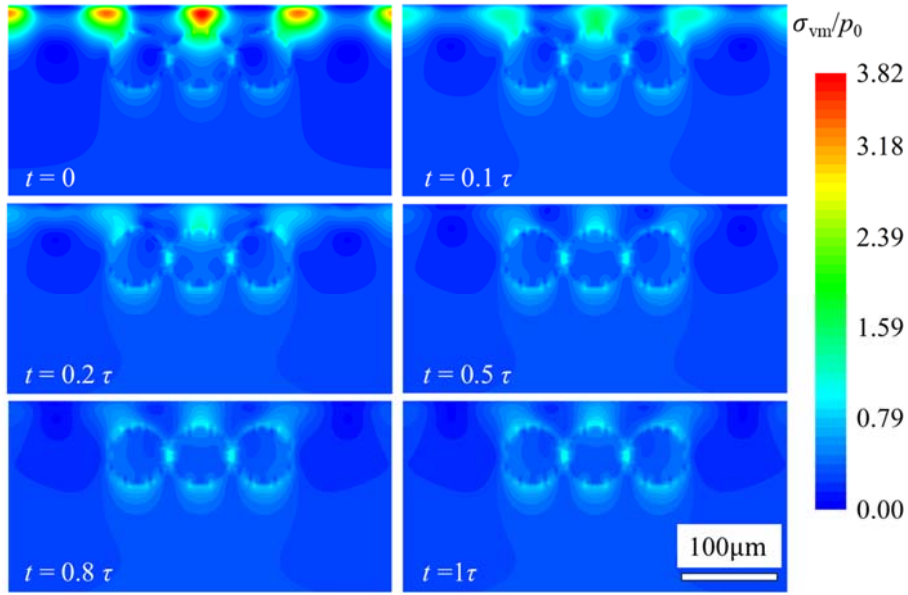

**Figure 10.** Contour maps of normalized von Mises stress at different time instants for a system with three inclusions under a stack pressure of $p_0$.

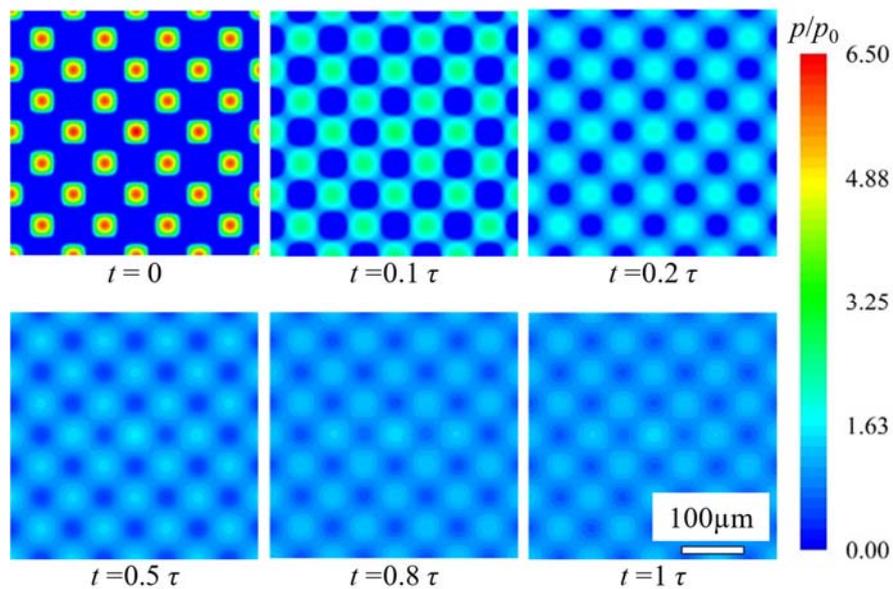

**Figure 11.** Contour maps of surface pressure at different time instants for a system with three inclusions under a stack pressure of $p_0$.

3.2.3 Effect of composite electrolyte composition in solid-state batteries performance

The contact behavior of interfaces in solid-state batteries is affected by several key features, such as the stack pressure, the material's inherent hardness, and the surface roughness. In this section, we consider the contact between a rough ceramics-in-polymer solid composite electrolyte (SCE) and an initially flat Li metal anode. More in detail, since we want to highlight the effect of SCE particle weight concentration, spatial distribution, and size on the overall SSB performance, we consider a solid composite electrolyte made of a purely elastic PEO polymer matrix with elasticity modulus $E_M = 82.8$ MPa [68, 73] and dispersed LLTO ceramic particles (inclusions) with approximate elastic modulus of $E_I = 150$ GPa. The final aim is to deepen the understanding of the material properties influencing solid-state battery performance, which is eventually characterized by the interface resistance $R_{int}$ [74, 75].

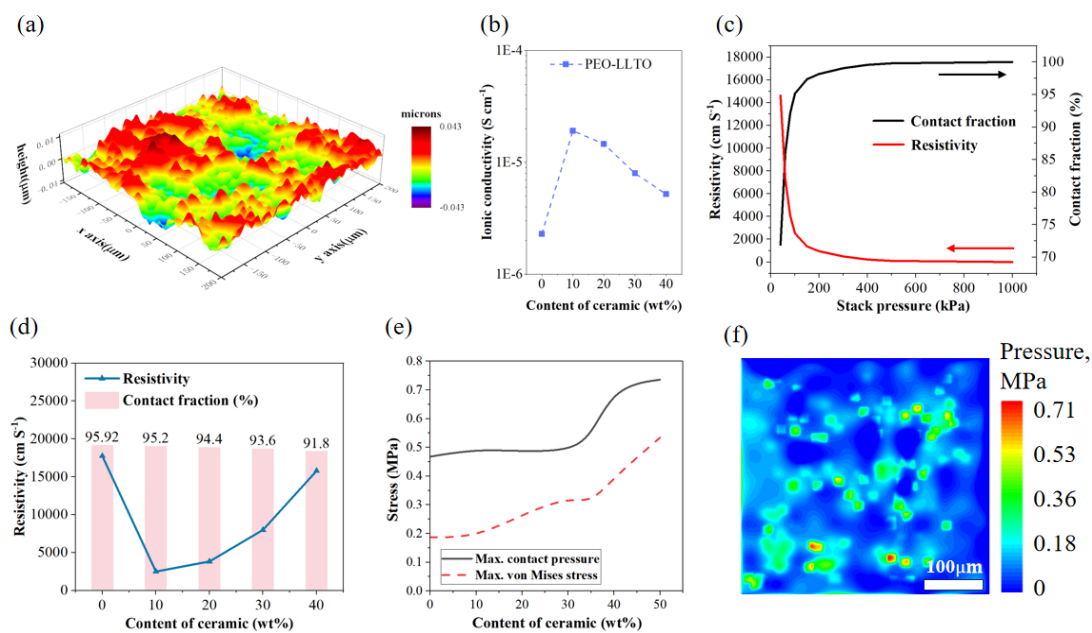

**Figure 12.** (a) Rough topography of the solid-state electrolyte surface, while the Li metal surface is initially smooth. (b) Ionic conductivity measurements for different amounts of LLTO in PEO matrix, from Refs. [76, 77]. (c) Interfacial resistivity and contact area fraction *vs.* stack pressure for SCE with LLTO particles (10 wt% content). (d) Interfacial contact area and interfacial resistivity with various LLTO contents under 100 kPa stack pressure. I Maximum contact pressure and maximum in-plane von Mises stress as functions of LLTO particles weight ratio (stack pressure is 100 kPa). (f) The contour of contact pressure for a SCE with 50 wt% LLTO particles (stack pressure is 100 kPa).

In the case of lithium metal anode and SSE, $R_{\text{int}}$ depends on the interphase film, charge transfer, and contact area [65, 72]; therefore, we have

$$R_{\text{int}} = R_{\text{int}}^0 + R_{\text{contact}} \qquad (10)$$

where $R^0{}_{\text{int}}$ represents a referenced resistance, which could either originates from interphase films, such as $Li_2CO_3$ [78], or be negligible, if the contact surfaces are sufficiently clean, devoid of oxidation films. It may also arise from other sources, such as grain boundaries or charge transfer processes [79]. For instance, charge transfer can often be disregarded if it is as minimal as $0.1\ \Omega\ \text{cm}^2$, like in the Li-LLTO interface, but becomes significant if it reaches higher values like $400\ \Omega\ \text{cm}^2$ at the Li-PEO interface [75, 80]. Moreover, in Eq. (10), $R_{\text{contact}} = ASR/F_c$ is the contact resistance, with the contact fraction $F_c$ being the ratio of the real contact area to the nominal contact area (notably, $F_c = 1$ when full contact occurs), and the areal specific resistance $ASR = 1/\sigma$ being a critical parameter in solid-state batteries (SSBs) directly linked to ionic conductivity $\sigma$ [79, 81, 82]. Notably, several studies [76, 77] have shown that at room temperature (25°C), the ionic conductivity of PEO-LLTO varies with particle content, as depicted in **Figure 12**(b).

Since $R_{\text{contact}}$, and in turn the interfacial resistance $R_{\text{int}}$, depends on the contact mechanics of the SCE and Li metal surface, a meaningful investigation of SSBs performance cannot ignore the role of the surface roughness of the contacting bodies. Indeed, **Figure 12**(a) shows the surface topography of a textured solid-state electrolyte (*i.e.*, the term $s_1$ in Eq. (1)), adapted from the similar study in [65], while the Li metal surface is assumed smooth at initial time (*i.e.*, $s_2(x, y, 0) = 0$ in Eq. (1)).

Given a particles concentration of 10 wt%, the effect of the stack pressure is reported in **Figure 12**(c), clearly suggesting a correlation between the interfacial resistivity and the contact fraction. On the other hand, **Figures 12**(d-e) present our findings on the overall effect of particles concentration, showing that a significant decrease in resistivity occurs at particle concentration of 10 wt% (which might even not be optimal [83]), while it does not correlate linearly with contact area fraction. Furthermore, the particle concentration leads to higher values of peak contact pressure and in-plane von Mises stress, which may lead to surface failure and defect propagation [83], eventually affecting the SSB durability. Moreover, higher particle concentration may increase the interfacial stress dispersion, as shown in **Figure 12**(f), which further leads to heterogeneous Li plating, eventually resulting in Li dendrites.

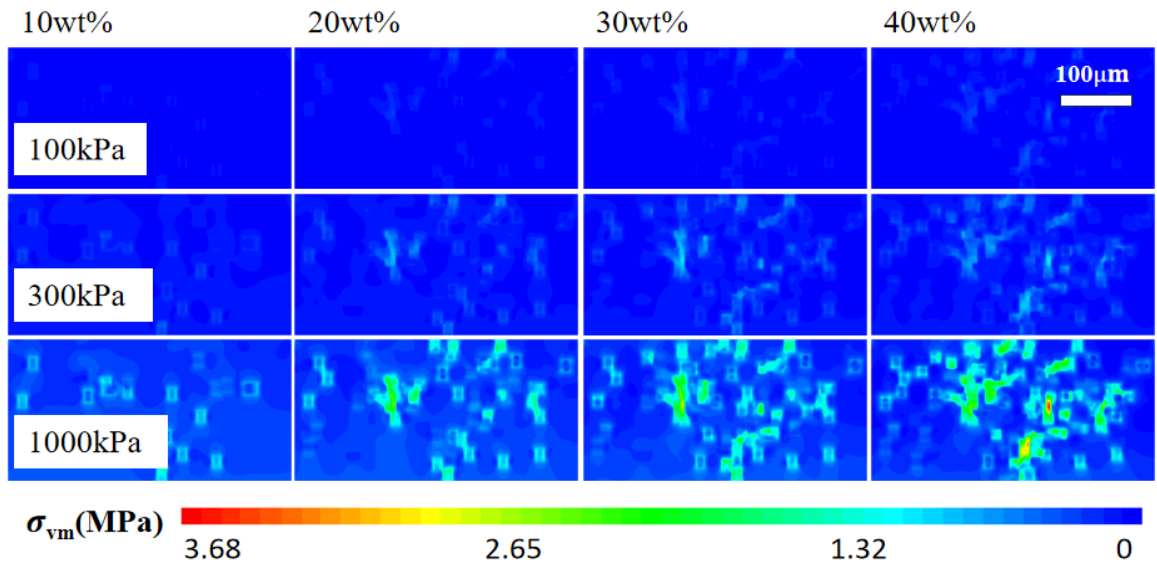

**Figure 13.** The von Mises stress maps for particle concentrations of 10 wt%, 20 wt%, 30 wt%, and 40 wt% under stacking pressures of 100, 300 and 1000 kPa.

Subsurface stress contours for particle concentrations of 10 wt%, 20 wt%, 30 wt%, and 40 wt% under different stack pressure values are shown in **Figure 13**, showing that the local stress is concentrated at the particle-matrix interface in systems under high pressures. Similarly, higher particle concentration leads to larger stress values, due to elastic interactions between inclusions.

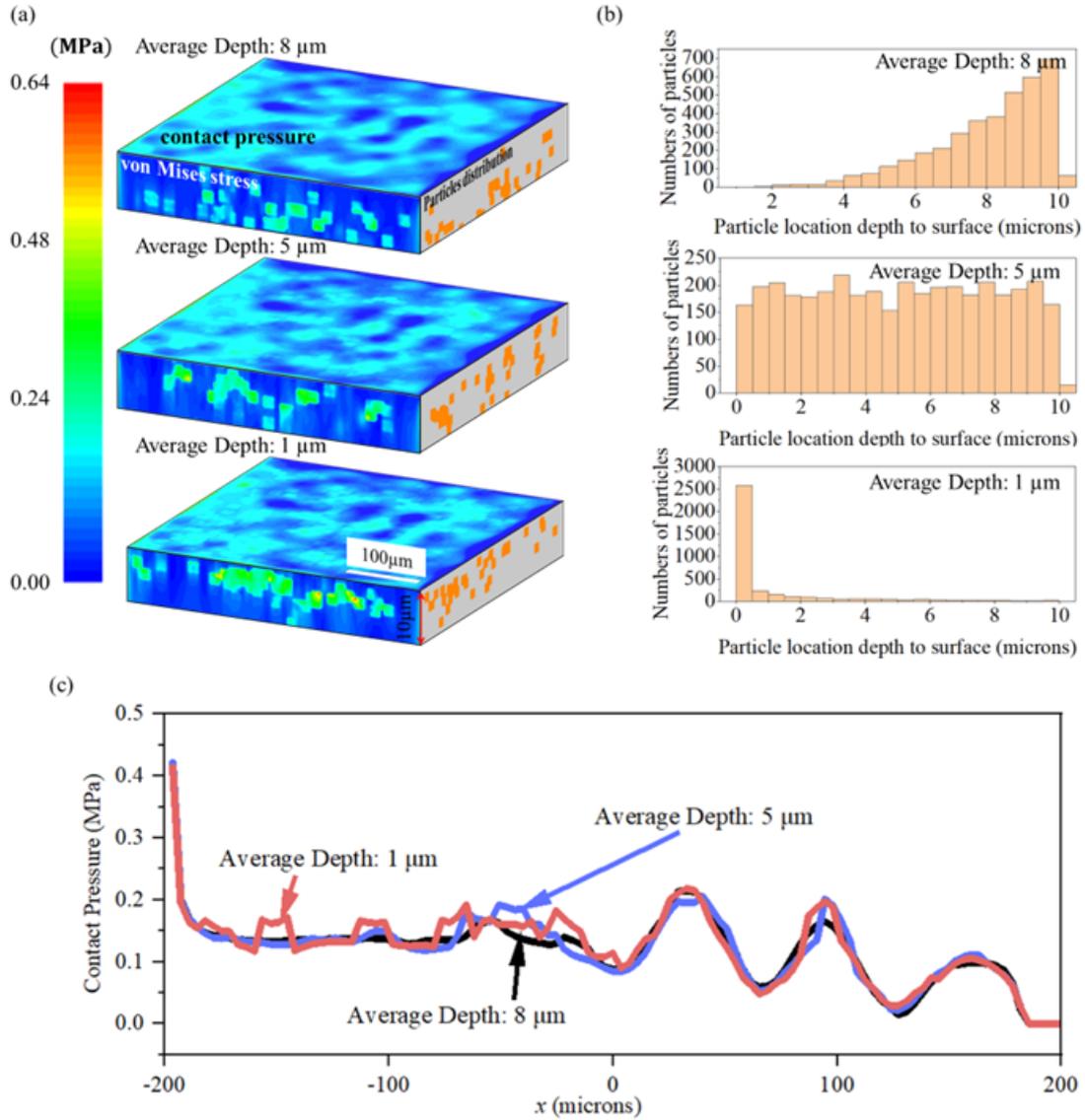

**Figure 14.** (a) The particle distribution (right face), the von Mises stress maps (front face) and the contact pressure maps (top face) for different average depths cubic particles, under a stacking pressure of 100 kPa. The particle size is 3.6 μm×3.6 μm ×2 μm. (b) Histograms of the number of particles as a function of the depth. (c) Contact pressure along the *x*-axis.

Recently, Zhang *et al.* [84] designed a solid-state polymer electrolytes with highly localized particle concentration, proving the importance to further investigate the influence of particle sizes and distributions on the overall interface performance. As shown in **Figure 14**, we simulated the contact under three different cubic particles distributions in the *z*-direction, which were generated using three sets of random numbers, maintaining the same particles number and a constant weight concentration of 40 wt%. The first set has an average depth of 8μm with a variance of 2.7. The second set has an average depth of 5μm with a variance of 8.3. The third set has an average depth of 1μm with a variance of 4.3. The average depth is calculated by $d_{Average} = \left(\sum d_i \times n_i\right)/n_{total}$, where $d_i$ is the depth of particle $i$, $n_i$ is the number of particles at location $d_i$, and $n_{total} = 3780$ is the total number of particles with. The top face of **Figure 14**(a) and the pressure profile in **Figure 14**(c) clearly indicate that the deeper the average depth is (i.e., the farther away from the surface the particles are located), the more homogeneous the surface contact becomes. This confirms that the design for solid-state electrolyte microstructures is crucial, as proposed for electrode's microstructure design [85].

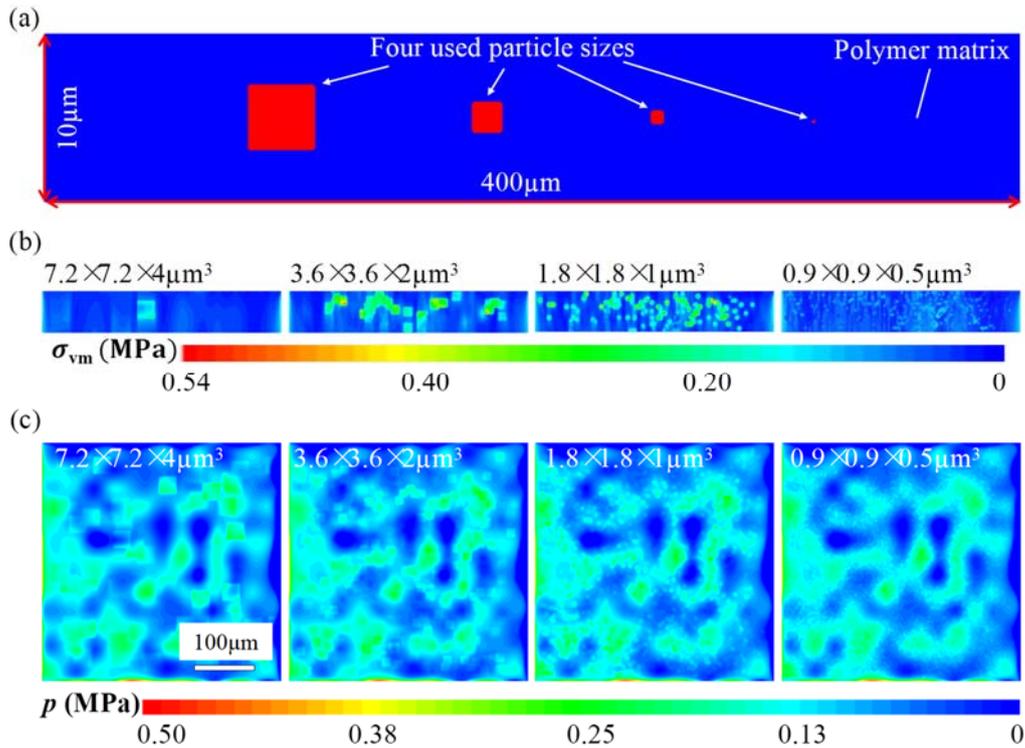

Figure 15. (a) Schematic diagram of comparison between particle sizes of four cases. (b) The von Mises stress maps for cubic particles with sizes of 7.2 μm×7.2 μm×4 μm, 3.6 μm×3.6 μm×2 μm, 1.8 μm×1.8 μm ×1 μm, and 0.9 μm×0.9 μm×0.5 μm under a stacking pressure of 100kPa. The average location depth of particles is kept to 5 μm. (c) Contour maps of surface contact pressure for cubic particles with sizes of 7.2 μm×7.2 μm×4 μm, 3.6 μm×3.6 μm ×2 μm, 1.8 μm×1.8 μm×1μ m, and 0.9 μm×0.9 μm×0.5 μm under a stacking pressure of 100kPa.

Moreover, since also the particle size is expected to significantly affect the contact behavior of the SCE interface, we simulated the contact scenarios for cubic particles with sizes of 7.2 μm×7.2 μm×4 μm, 3.6 μm×3.6 μm ×2 μm, 1.8 μm×1.8 μm ×1 μm, and 0.9 μm×0.9 μm×0.5 μm, with a particles' average depth of 5 μm (*i.e.*, conditions corresponding to the second set of **Figure 14**). Their specific dimensions and the von Mises stress maps are shown in **Figure 15**(a) and **Figure 15**(b), respectively. As shown in **Figure 15**(c), we found that there was no significant change in the peak of contact pressure, although smaller particles lead to more homogeneous contact pressure distribution.

The presented results provide useful insight on the performance of composite solid-state electrolytes at both micro- and macro-scale. At the micro-scale, i.e. close to the particle-matrix interface, the proposed model allows to investigate the stress concentration, thus helping in predicting the initiation and propagation of fractures and cracks in weak points. At the macro-scale, our model predicts the surface stress distribution and the real contact area, thus establishing a correlation between localized mechanical behaviors and the overall structural integrity and functionality of solid-state electrolytes and offering potential paths of SSB optimization and predictive modeling in [86].

3.2.4 Time-dependent Li-SSE contact and Solid-State Battery performance

During the processes of charging or discharging, the contact interface between the lithium and the solid electrolyte (Li-SSE) is subject to both electrochemical reactions and chemomechanical deformations. Notably, these phenomena are time-dependent, and can significantly alter the interface's mechanical response over time and, in turn, exert a substantial impact on SSB performance. Consequently, understanding the complex interactions at the Li-SSE interface is of utmost importance in the optimization process of electrochemical storage technologies.

Here, we consider a solid composite electrolyte (SCE) made of a PEO polymer matrix and dispersed LLTO ceramic particles (inclusions) with approximate elastic modulus of $E_I = 150$ GP. The PEO is now considered as a viscoelastic material, in which the complex viscoelastic modulus $G(\omega)$ (shear modulus) was measured by experiments in [87], where the real portion, Re[$G(\omega)$], and the imaginary portion, Im[$G(\omega)$], are shown in **Figure 16(a)**. Using the target equation $G(\omega) = 1/\left[\frac{1}{G_0} - \sum_{k=1}^{N} S_k \frac{i\omega\tau_k}{1+i\omega\tau_k}\right]$ to approach the experimental data in **Figure 16**(a), we have obtained parameters of $G_0 = 2(1+\nu)E_M = 215$MPa (consistent with the PEO's elasticity modulus $E_M = 82.8$ MPa [68, 73] used in Sec 3.2.3), and $S_k$ (listed in **Table 1**), by fitting Re[$G(\omega)$] and Im[$E(\omega)$], as plotted in **Figure 16**(a), based on the non-linear fitting method introduced by Zhang *et al.* [88] with geometric progression $\tau_{k+1}/\tau_k = 2e$ and $\tau_1 = 0.01$, and ten relaxation times, *i.e.* N=10.

Table 1. Fitted parameters $S_k$ of the PEO material.

| | |
|---|---|
| $S_1$ | 7.19×10⁻³+1.85×10⁻²i |
| $S_2$ | 0 |
| $S_3$ | 9.87×10⁻⁵-1.45×10⁻³i |
| $S_4$ | -4.47×10⁻⁴-7×10⁻⁴i |
| $S_5$ | -6.95×10⁻⁴-4.78×10⁻⁴i |
| $S_6$ | 6.11×10⁻²-8.83×10⁻²i |
| $S_7$ | -3.74+2.25i |
| $S_8$ | 19.6-9.93i |
| $S_9$ | -4.46+2.59i |
| $S_{10}$ | -11.4+5.18i |

Following the procedure detailed in Ref. [65], we simulated the plating rate as a function of the local current, as well as the elasto-plastic behavior of lithium metal anode and it's time-dependent creep behavior. The SCE behavior is described by the present model, taking into account for different ceramic particles concentrations. More in detail, for a 2.5-hour plating process, we conducted 311 contact simulations, 310 creep computations, and 310 mechanical-electrochemistry calculations, focusing on a single asperity contact, whose initial shape is shown in **Figure 16**(b).

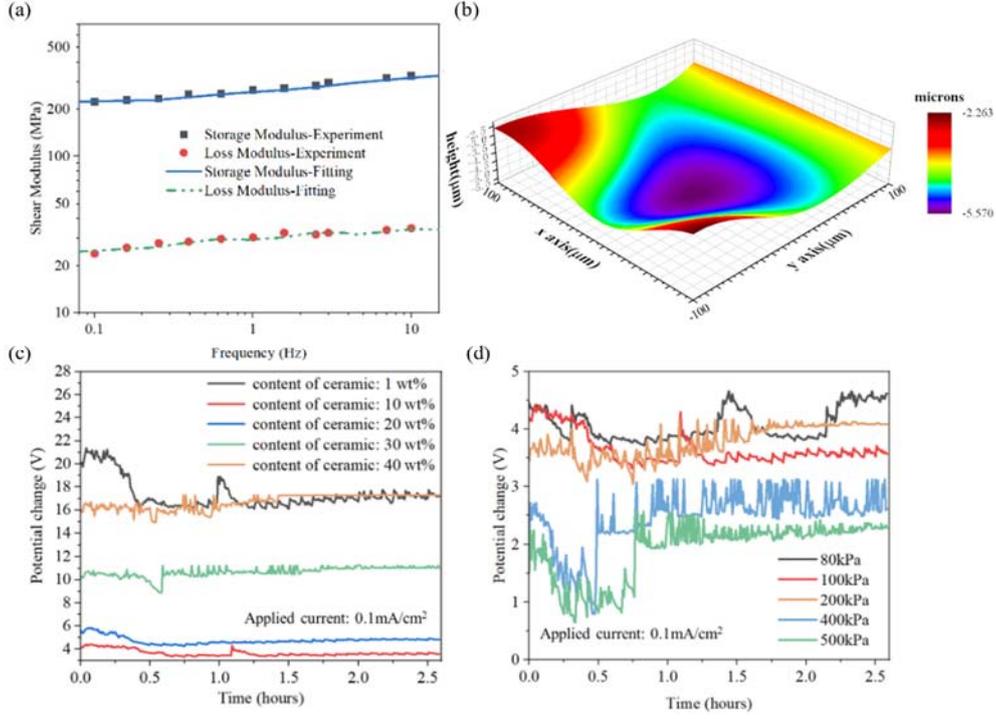

**Figure 16.** (a) Measured data and their fitting curves of the complex viscoelastic shear modulus with its real portion (storage modulus) and imaginary portion (loss modulus), (b) Topographic map of single asperity contact time $t = 0$. Interface potential evolution over plating time for (c) different ceramic particles concentrations (stack pressure is 100 kPa) and (d) different stack pressure (particles concentration is 10 wt%). Calculations assume electrode current density $I = 0.1 \text{mA/cm}^2$.

The variation in interfacial potential during a 2.5-hour electroplating process is shown in **Figures 16**(c-d) for different ceramic particles concentrations and stack pressure, respectively. The results indicate very low interfacial potential for a ceramic content of 10 wt%, in agreement with the results presented in **Figures 12**(b, d) for ionic conductivity and interfacial resistance; moreover, the potential is also almost constant during the whole simulation (see **Figure 16**(c)), suggesting that 10 wt% is the optimal ceramic content for minimizing potential fluctuations and enhancing interface stability. Finally, as suggested in Ref. [65], increasing the stack pressure leads to higher contact fraction values $F_c$ and, in turn, through Eq. (10), to lower interfacial resistance and potential (as shown in **Figure 16**(d)).

## 4. CONCLUSIONS

A numerical method based on the hybrid element method (HEM) has been proposed to model the multi-scale three-dimensional time-dependent contact between Li and solid composite electrolytes (SCEs). The model has been validated by contrasting the computational results of an elastic point contact analysis with a single spherical inclusion degenerated from the presented method and those by Zhou *et al.* [71]. Furthermore, we compare outcomes for a frictionless rigid spherical indenter sliding over a viscoelastic half-space with an encapsulated elastic spherical inhomogeneity, with parallel work by Koumi *et al.* [59] . Contact pressure distribution and subsurface stresses have been analyzed in the presence of pure viscoelastic matrix, a single and three spherical heterogeneities. The effect of the reinforcement on the polymer's modulus and on its conductivity has also been presented. The main results are summarized below:

1. Incorporating ceramic particles into a polymer matrix has been shown to influence both surface pressure distribution and subsurface stress distribution. Moreover, the addition of more particles has been observed to elevate the maximum surface pressure and stress levels.

2. Viscoelasticity has been observed to mitigate the impact of heightened surface pressure and subsurface stress resulting from heterogeneity, without changing their spatial localization.

3. The surface contact area can be calculated for different types of polymer matrices, ceramic particle numbers, distributions, and various stacking pressures. Additionally, the interfacial resistance of the solid-state batteries can be determined by integrating the corresponding conductivity.

4. Maintaining a constant average location depth of particle distributions, the incorporation of smaller ceramic particles results in more homogeneous distributions of the contact pressures, while shows a negligible impact on the peak contact pressure. Conversely, keeping a fixed particle size, the deeper the average depth of particle distribution (thereby the further the distance of particles from the surface), the more homogeneous the contact pressure.

5. Different ceramic content, applied current and stacking pressure will all affect the change of interface potential. Under certain conditions (10 wt% ceramic content, 100 kPa stacking pressure and a certain electrode current density), the change of interface potential is the smallest, showing the best interface stability. These findings emphasize the importance of understanding the interface dynamics of solid-state batteries for optimizing battery performance and suggest the possibility of improving battery performance by adjusting mechanical properties and operating parameters.

6. To optimize the design of solid-state electrolytes (SSEs), it is necessary to incorporate a lithium ion conduction model. While this study does not currently include such a model, integrating it in future work would provide valuable insights into the ion transport mechanisms and could significantly improve the optimization of SSEs.

ACKNOWLEDGMENTS
The authors would like to acknowledge support by the National Natural Science Foundation of China (12102085), and the Postdoctoral Science Foundation of China (2023M730504). This work was partly supported by the European Union – NextGenerationEU through the Italian Ministry of University and Research under the program PRIN2022 (Projects of Relevant National Interest) grant nr. 2022SJ8HTC - ELectroactive gripper For mIcro-object maNipulation (ELFIN); PRIN2022 PNRR (Projects of Relevant National Interest) grant nr. P2022MAZHX - TRibological modellIng for sustainaBle design Of induStrial friCtiOnal inteRfacEs (TRIBOSCORE). The opinions expressed are those of the authors only and should not be considered as representative of the European Union or the European Commission's official position.